---------------------------------------------------------------------------
############################## .TEX FILE ###################################
---------------------------------------------------------------------------

\documentclass[11pt]{article}    
\usepackage{amssymb}
\def\u{\leavevmode\hbox{\normalsize1\kern-4.2pt\large1}}
\def\N{\mbox{\tiny $N$}}
\def\d{\mbox{\tiny $2N$}}
\def\-{\mbox{\tiny $-$}}
\def\+{\mbox{\tiny $+$}}
\def\sdag{\mbox{\tiny $\dag$}}
\def\salpha{\mbox{$\frac{\alpha}{2}$}}
\def\f{\mbox{$\frac{1}{2}$}}
\def\L{\mbox{\tiny $L$}}
\def\R{\mbox{\tiny $R$}}
\def\Y{\mbox{\tiny $Y$}}
\def\0{\mbox{\tiny $0$}}
\def\1{\mbox{\tiny $1$}}
\def\2{\mbox{\tiny $2$}}
\def\3{\mbox{\tiny $3$}}
\def\4{\mbox{\tiny $4$}}
\def\8{\mbox{\tiny $8$}}
\def\6{\mbox{\tiny $16$}}
\def\t{\mbox{\tiny $00$}}
\def\w{\mbox{\tiny $01$}}
\def\od{\mbox{\footnotesize $2 \times 2$}} 
\def\oq{\mbox{\footnotesize $4 \times 4$}} 
\def\on{\mbox{\footnotesize $N \times N$}} 
\def\ond{\mbox{\footnotesize $2N \times 2N$}} 
\def\stx{\mbox{$\frac{\theta_x}{2}$}}

\def\svx{\mbox{$\frac{\varphi_x}{2}$}}

\def\b{\mbox{\tiny $/$}}

\begin{document}

%%%%%%%%%%%%%%%%%%%%%%%%%%%%%%%% PAPER  %%%%%%%%%%%%%%%%%%%%%%%%%%%%%%%%%%%%

\title{{\Huge \sf Quaternionic Groups in Physics:\\
                   A Panoramic Review}}
\author{
{\Large \sf Stefano De Leo$^{(a,b)}$ {\sf and} Gisele C.~Ducati$^{(a,b)}$}\\ \\
{\small $^{(a)}${\em Departamento de Matem\'atica Aplicada, UNICAMP}}\\
{\small  CP 6065, 13081-970  Campinas (SP) Brasil}\\
{\small  \tt deleo/ducati@ime.unicamp.br}\\
{\small  $^{(b)}${\em Dipartimento di Fisica, Universit\'a degli Studi 
di Lecce}}\\
{\small  via Arnesano, CP 193, 73100 Lecce, Italia}\\
{\small  \tt deleos@le.infn.it}\\
{\small  $^{(c)}${\em Departamento de Matem\'atica, UFPR}}\\
{\small  CP 19081, 81531-970  Curitiba (PR) Brasil}\\
{\small  \tt ducati@mat.ufpr.br}
} 
\date{December, 1998}
\maketitle

%%%%%%%%%%%%%%%%%%%%%%%%%%%%%%%%%%%%%%%%%%%%%%%%%%%%%%%%%%%%%%%%%%%%%%%%%%%%%%%
%                                ABSTRACT
%%%%%%%%%%%%%%%%%%%%%%%%%%%%%%%%%%%%%%%%%%%%%%%%%%%%%%%%%%%%%%%%%%%%%%%%%%%%%%%

\begin{abstract}

Due to the non-commutative nature of quaternions we introduce 
the concept of left and right action for quaternionic numbers.
This gives the opportunity to manipulate 
appropriately the $\mathbb{H}$-field. The standard problems arising in the 
definitions of transpose, determinant and trace for quaternionic matrices are
overcome. We investigate the possibility to formulate a 
{\em new approach} to Quaternionic Group Theory. Our aim is to highlight the 
possibility of looking at new quaternionic groups by the use of left and 
right operators as fundamental step toward a clear and complete 
discussion of Unification Theories in Physics.

%{\bf Key words:} 

%{\bf 1996 PACS numbers:}

%{\bf 1991 Mathematics Subject Classification:} 

\end{abstract}

%%%%%%%%%%%%%%%%%%%%%%%%%%%%%%%%%%%%%%%%%%%%%%%%%%%%%%%%%%%%%%%%%%%%%%%%%%%%%%%
%                                SECTION I
%%%%%%%%%%%%%%%%%%%%%%%%%%%%%%%%%%%%%%%%%%%%%%%%%%%%%%%%%%%%%%%%%%%%%%%%%%%%%%%

\section{Introduction}
\label{s1}

Complex numbers have played a dual role in Physics, first as a technical tool
in resolving differential equation (classical optics) or via the theory
of analytic functions for performing real integrations, summing series, etc.;
secondly in a more essential way in the development of Quantum Mechanics and 
later Field Theory. With quaternions, for the first type of 
application, i.e. as a means to simplify calculations, we can quote the 
original work of Hamilton~\cite{HAM}, 
but this only because of the late development of 
vector algebra by Gibbs and Heaviside~\cite{GH}. Even Maxwell used 
quaternions as a 
tool in his calculations, e.g. in the 
{\sl Treatise of Electricity and Magnetism}~\cite{MAX} 
where we find the $\nabla$-operator expressed by the three quaternionic 
imaginary units.

Notwithstanding the 
Hamilton's conviction that quaternions would soon play a role 
comparable to, if not greater than, that of complex numbers the use of 
quaternions in Physics was very limited~\cite{GUR}. Nevertheless, in the 
last decades, we 
find a renewed interest in the application of non-commutative fields in 
Mathematics and Physics. In Physics, we quote quaternionic versions of
%Gauge Theories~\cite{FIN1,HOR,QET,GUT}, 
Gauge Theories~\cite{FIN1}-\cite{GUT}, 
Quantum Mechanics and 
%Fields~\cite{FIN2,ADL1,ADL2,DIRAC,DKP,DR1},
Fields~\cite{FIN2}-\cite{DR1},
Special Relativity~\cite{SR}. In Mathematics, we find applications of 
quaternions for Tensor Products~\cite{RH,DR2}, 
Group Representations~\cite{ADL3}. 

In this paper we aim to give a {\em new} panoramic review of quaternionic  
groups. We use the adjective ``new'' since the elements of our matrices   
will not be simple quaternions but {\em left and right operators}, 
originally introduced with the name of ``barred operators''~\cite{EVEN}.

In Physics, particularly Quantum Mechanics, we are accustomed to 
distinguishing between ``states'' and ``operators''. Even when the operators
are represented by numerical matrices, the squared form of operators 
distinguishes them from the column structure of the spinors states. Only for 
one-component fields and operators is there potential confusion. In 
extending Quantum Mechanics defined over the complex field to quaternions
it has almost always been assumed that matrix operators
contain elements which are ``numbers'' indistinguishable from those of the 
state vectors. {\em This is an unjustified limitation}. In fact, 
(non-commutative) hyper-complex theories require left/right operators.

This paper is organized as follows: In section~\ref{s2}, we introduce 
the quaternionic algebras. In section~\ref{s3}, we show that 
the non-commutative nature of the quaternionic field suggests 
the use of left/right operators.  In section~\ref{s4}, 
we find the appropriate 
definitions of transpose, trace and determinant for quaternionic
matrices. Such a section contains the {\em new} classification of
quaternionic groups. In section~\ref{s5}, we present some applications 
of left/right operators in Physics. Our conclusions are drawn in the 
final section.

%%%%%%%%%%%%%%%%%%%%%%%%%%%%%%%%%%%%%%%%%%%%%%%%%%%%%%%%%%%%%%%%%%%%%%%%%%%%%%%
%                                SECTION II
%%%%%%%%%%%%%%%%%%%%%%%%%%%%%%%%%%%%%%%%%%%%%%%%%%%%%%%%%%%%%%%%%%%%%%%%%%%%%%%

\section{Quaternionic States}
\label{s2}

Complex numbers can be constructed from the real numbers by introducing
a quantity $i$ whose square is $\- 1$:
\[ c = r_{\1} + i r_{\2} \quad \quad (r_{\1 , \2} \in {\mathbb{R}}) \; .\]
Likewise, we can construct the quaternions from the complex numbers in exactly
the same way by introducing another quantity $j$ whose square is $\- 1$,
\[ q = c_{\1} + j c_{\2} \quad \quad (c_{\1 ,\2} \in \mathbb {C})\; ,\]
and which anti-commutes with $i$ 
\[ij=\- ji=k~.\]

In introducing the quaternionic algebra, let us follow the conceptual 
approach of Hamilton. In 1843, the Irish mathematician attempted to generalize
the complex field in order to describe the rotations in the three-dimensional 
space. He began by looking for numbers of the form 
\[x+iy+jz~,\] 
with $i^{\2}=j^{\2}=\- 1$. Hamilton's hope was to do for three-dimensional 
space what complex numbers do   
for the plane. Influenced by the existence of a complex number norm
\[ c^* c = (\mbox{Re} \, c)^{\2} +  (\mbox{Im} \, c)^{\2}~,\]
when he looked at its generalization 
\[ (x-iy-jz) (x+iy+jz) = 
    x^{\2}+y^{\2}+z^{\2} - (ij+ji) yz~,\]
to obtain a real number, he had to adopt  the anti-commutative law of 
multiplication for the imaginary units. Nevertheless, 
with only two imaginary 
units we have no chance of constructing a new 
numerical field, because assuming
\[
ij=\alpha + i \beta + j \gamma
~~~~~~~(\alpha , \beta , \gamma \in \mathbb{R})~,
\]
and using the multiplication associativity, $ \- i(ij) = \- i^{\2} j = j $, 
we find
\[ 
j = \beta - i \alpha - ij \gamma = i \alpha - \beta - 
\left( \alpha + i \beta + j \gamma \right) \gamma~,\]
which implies 
\[\alpha = \beta =0~~~~~\mbox{and}~~~~~\gamma^{\2}=\- 1~.
\]
Thus, we must introduce a third imaginary unit $k\neq i,j$, with
\[ k=ij= \- ji~. \] 
The $\mathbb{H}$-field is therefore characterized by three imaginary 
units $i$, $j$, $k$ which satisfy the following multiplication rules
\begin{equation}
i^{\2}=j^{\2}=k^{\2}=ijk= \- 1~.
\end{equation}
Numbers of the form
\begin{equation}
\label{q}
q=x_{\0}+ ix+jy+kz~~~~~~~~(~x_{\0},x,y,z \in {\mathbb{R}}~)~,
\end{equation}
are called (real) {\em quaternions}. They 
are added, subtracted and multiplied according to the usual laws of 
arithmetic, except for the commutative law of multiplication.

Similarly to  rotations in a plane that can be concisely expressed by complex 
number, a
rotation about an axis passing through the origin and parallel to a given 
unitary vector $\hat{u}\equiv(u_x,u_y,u_z)$ by an angle $\alpha$ can be 
obtained taking the following {\em quaternionic} transformation
\[ \exp\left( \salpha \, \vec{h}\cdot \vec{u} \right)
\, \vec{h}\cdot \vec{r} \, 
\exp \left(\- \salpha \, \vec{h}\cdot \vec{u} \right) ~ , \]
where 
\[
\vec{h}\equiv (i,j,k)~~~~~\mbox{and}~~~~~\vec{r} \equiv (x,y,z)~. 
\]
In section~\ref{s5}, 
we shall see how the quaternionic number $q$ in Eq.~(\ref{q}),
with the identification $x_{\0}\equiv ct$, can be used to formulate a 
one-dimensional version of the Lorentz group~\cite{SR}. We obtain 
the natural 
generalization of Hamilton's idea
\begin{center}
complex/plane~~~~~$\rightarrow$~~~~~pure imaginary 
quaternions/space~~~~~$\rightarrow$~~~~~quaternions/space-time~,
\end{center}
completing the unification of algebra and geometry.

Let us now consider the (full) conjugate of $q$ 
\begin{equation}
\label{qb} 
q^{\sdag} = x_{\0} - i x - j y - k z~.
\end{equation}
We observe that $q^{\sdag}q$ and $qq^{\sdag}$ are both equal to the 
real number
\[ N(q) = x_{\0}^{\2} + x^{\2} + y^{\2} + z^{\2}~,\]
which is called the norm of $q$. When $q\neq 0$, we can define
\[ q^{\- \1} = q^{\sdag} / N(q)~,\]
so quaternions form a zero-division ring.

An important difference between quaternionic and complex numbers is related 
to the definition of the conjugation operation. Whereas with complex 
numbers we can define only one type of conjugation
\[ i \rightarrow \- i~,\]
working with quaternionic numbers we can introduce 
different conjugation operations.  
Indeed, with three imaginary units we
have the possibility to define besides the standard conjugation~(\ref{qb}), 
 the six new operations
\begin{eqnarray*}
(i,j,k) & ~\rightarrow ~& 
(\- i,\+ j,\+ k)~,~(\+ i,\- j,\+ k)~,~(\+ i,\+ j,\- k)~;\\
(i,j,k) & ~\rightarrow ~& 
(\+ i,\- j,\- k)~,~(\- i,\+ j,\- k)~,~(\- i,\- j,\+ k)~.
\end{eqnarray*}
These last six conjugations can be concisely represented by $q$ and 
$q^{\sdag}$ as follows
\begin{eqnarray*}
q & ~\rightarrow ~& \- iq^{\sdag}i~,~\- jq^{\sdag}j~,~\- kq^{\sdag}k~,\\
q & ~\rightarrow ~& \- iqi~~,~\- jqj~~,~\- kqk~~.
\end{eqnarray*}
It could seem that the only independent conjugation be represented by 
$q^{\sdag}$. Nevertheless, $q^{\sdag}$ can also be expressed 
in terms of  $q$, in fact 
\begin{equation}
\label{qbar}
q^{\sdag} = \- \f \; (q+iqi+jqj+kqk)~.
\end{equation}

We conclude this section by introducing a compact notation to represent 
quaternionic states. Let
\begin{equation}
\label{h}
h^\mu \equiv \left( \, 1 \, , \, \- \vec{h} \, \right)~~~~~\mbox{and}~~~~~
h_\mu \equiv \left( \, 1 \, , \, \vec{h} \, \right)
\end{equation}
denote the $\mathbb{H}$-field generators and  
\begin{equation}
\label{x}
x^\mu \equiv \left( \, x_{\0} \, , \, \vec{x} \, \right)~~~~~\mbox{and}~~~~~
x_\mu \equiv \left( \, x_{\0} \, , \, \- \vec{x} \, \right)
\end{equation}
be contravariant and covariant real quadrivectors,
\[ g^{\mu \nu} = ( \+ , \- , \- , \- )~. \]
Quaternionic states will be 
written in terms of~(\ref{h},\ref{x}) as follows
\[ 
q = x^\mu h_\mu = x_{\mu} h^{\mu} \equiv x_{\0} + \vec{h} \cdot \vec{x}~,
\] 
and consequently the (full) quaternionic conjugate, $q^{\sdag}$ will read
\[
q^{\sdag} = x^\mu h^\mu = x_{\mu} h_{\mu}
\equiv x_{\0} - \vec{h} \cdot \vec{x}~.
\]
From now on, Greek letters will be run from \mbox{\small $0$} to 
\mbox{\small $3$}.

%%%%%%%%%%%%%%%%%%%%%%%%%%%%%%%%%%%%%%%%%%%%%%%%%%%%%%%%%%%%%%%%%%%%%%%%%%%%%%%
%                                SECTION III
%%%%%%%%%%%%%%%%%%%%%%%%%%%%%%%%%%%%%%%%%%%%%%%%%%%%%%%%%%%%%%%%%%%%%%%%%%%%%%%

\section{Quaternionic Left/Right Operators}
\label{s3}

Due to the non-commutative nature of quaternions we must distinguish between 
$q \vec{h}$ and $\vec{h}q$. Thus,  it is appropriate to consider left and
right-actions for our imaginary units $i$, $j$ and $k$. 
Let us define the operators
\begin{equation}
L_{\mu} \equiv \left( \, \u \, , \, \vec{L} \, \right) 
\in {\mathbb{H}}^{\L}~,~~~~~
\vec{L} = \left(  L_i , L_j , L_k  \right)~,
\end{equation} 
and
\begin{equation}
R_{\mu} \equiv \left( \, \u \, , \, \vec{R} \, \right) 
\in {\mathbb{H}}^{\R}~,~~~~~
\vec{R} = \left(  R_i , R_j , R_k  \right)~,
\end{equation} 
which act on quaternionic states in the following way
\begin{equation}
L_{\mu} : ~\mathbb{H} \rightarrow \mathbb{H}~,~~~~~     
L_{\mu} q = h_\mu q \in \mathbb{H}~,
\end{equation}
and
\begin{equation}
R_{\mu} : ~\mathbb{H} \rightarrow \mathbb{H}~,~~~~~     
R_{\mu} q = q \, h_\mu \in \mathbb{H}~.
\end{equation}
The algebra of left/right generators can be concisely expressed by 
\[ 
L_i^{\2} = L_j^{\2} =L_k^{\2} =L_i L_j L_k =
R_i^{\2} = R_j^{\2} =R_k^{\2} =R_k R_j R_i =
\- \u~,
 \]
and by the commutation relations
\[ 
\left[ \, L_{i,j,k} \, , \, R_{i,j,k} \, \right] = 0~.
\]

In this section we will discuss three different types of operators. 
Operators $\mathbb{H}$-linear, $\mathbb{C}$-linear, $\mathbb{R}$-linear
from the right. For simplicity of notation we introduce
\[ 
{\cal O}_{\mathbb{X}} : ~\mathbb{H} \rightarrow \mathbb{H}~, 
\] 
to represent quaternionic operators right-linear on the 
$\mathbb{X}$-field. Operators which act only from the left,
\[ 
{\cal O}_{\mathbb{H}} = a^{\mu} L_{\mu} \in \mathbb{H}^{\L}~,~~~~~
a^{\mu} \in \mathbb{R}^{\4}~,
\]
are obviously $\mathbb{H}$-linear from the right
\[
L_{\mu} ( q \lambda ) = \left( L_{\mu}  q \right) \lambda~,~~~~~
\lambda \in  \mathbb{H}~,
\]
and $\mathbb{R}$-linear from the left
\[
L_{\mu} ( \rho q ) =  \rho \left( L_{\mu}  q \right) ~,~~~~~
\rho \in  \mathbb{R}~.
\]
Let us now consider the sixteen generators
\[ M_{\mu \nu} \equiv L_{\mu} \otimes R_{\nu}~.\]
Due to left and right actions of the imaginary units $i$, $j$, $k$  the
corresponding operator
\[ 
{\cal O}_{\mathbb{R}} = 
a^{\mu \nu} M_{\mu \nu} \equiv 
a^{\mu \nu} L_{\mu} \otimes R_{\nu}
\in \mathbb{H}^{\L} \otimes \mathbb{H}^{\R}~, ~~~~~
a^{\mu \nu} \in \mathbb{R}^{\6}~,
\]
are restricted to be $\mathbb{R}$-linear 
\[ 
M_{\mu \nu } ( q \rho ) = 
\left( M_{\mu \nu }  q \right) \rho =
\rho \left( M_{\mu \nu }  q \right)~,~~~~~
\rho \in  \mathbb{R}~.
\]
Finally, considering the right action to the only $i$-complex imaginary unit
\[
M_{\mu n} \equiv L_{\mu} \otimes R_{n}~,~~~~~
\mbox{\tiny $n=1,2$}~,
\]
we obtain right $\mathbb{C}$-linear operators
\[ 
{\cal O}_{\mathbb{C}} = 
a^{\mu n} M_{\mu n} \equiv 
a^{\mu n} L_{\mu} \otimes R_{n}
\in \mathbb{H}^{\L} \otimes \mathbb{C}^R
\subset \mathbb{H}^{\L} \otimes \mathbb{H}^{\R}~, ~~~~~
a^{\mu n} \in \mathbb{R}^{\8}~,
\]
in fact
\[ 
M_{\mu n } ( q \zeta ) = \left( M_{\mu n }  q \right) \zeta~,~~~~~
\zeta \in  \mathbb{C}~.
\]
The classification of right $\mathbb{H}$/$\mathbb{C}$/$\mathbb{R}$-linear
can be summarized by 
\[
{\cal O}_{\mathbb{H}} \subset {\cal O}_{\mathbb{C}} \subset 
{\cal O}_{\mathbb{R}}~,
\]
note that all these operators are obviously 
$\mathbb{R}$-linear from the left.

Let us now analyze the product of two right $\mathbb{R}$-linear operators 
\[ 
{\cal O}_{\mathbb{R}}^a = a^{\mu \nu} M_{\mu \nu}~,~~~
{\cal O}_{\mathbb{R}}^b = b^{\tau \sigma} M_{\tau \sigma}~,
\]
in terms of left/right quaternionic generators is given by
\[ 
{\cal O}_{\mathbb{R}}^a \, {\cal O}_{\mathbb{R}}^b =  
a^{\mu \nu} b^{\tau \sigma} L_{\mu} L_{\tau} \otimes R_{\sigma} R_{\nu}~.
\]
From such a relation we can immediately obtain the product of 
right $\mathbb{C}$-linear and $\mathbb{R}$-linear operators. 
The {\em full} conjugation operation for left/right operators is defined by 
a simultaneous change in the sign of left/right quaternionic 
imaginary units, i.e.
\[ 
L_{\mu}^{\sdag} = L^{\mu} = g^{\mu \nu} L_{\nu}~~~~~\mbox{and}~~~~~
R_{\mu}^{\sdag} = R^{\mu} = g^{\mu \nu} R_{\nu}~. 
\]
Thus, 
\begin{equation}
\label{lrd}
{\cal O}_{\mathbb{R}}^{\sdag} = a^{\mu \nu} 
L_{\mu}^{\sdag} \otimes R_{\nu}^{\sdag} =  
a^{\mu \nu} L^{\mu} \otimes R^{\nu}~~~
\in \mathbb{H}^{\L} \otimes \mathbb{H}^{\R}~.
\end{equation}
For operators product conjugations we have 
\begin{eqnarray*}
\left( {\cal O}_{\mathbb{R}}^a \, {\cal O}_{\mathbb{R}}^b \right)^{\sdag} 
& = &  
a^{\mu \nu} b^{\tau \sigma} 
\left( L_{\mu} L_{\tau} \otimes R_{\sigma} R_{\nu} \right)^{\sdag}\\
& = &  
a^{\mu \nu} b^{\tau \sigma} 
\left( L_{\mu} L_{\tau} \right)^{\sdag} 
\otimes \left( R_{\sigma} R_{\nu} \right)^{\sdag}\\
& = & 
a^{\mu \nu} b^{\tau \sigma} 
 L_{\tau}^{\sdag} L_{\mu}^{\sdag} \otimes R_{\nu}^{\sdag} 
R_{\sigma}^{\sdag} \\
& = & 
{\cal O}_{\mathbb{R}}^{b \, \sdag} \,  
{\cal O}_{\mathbb{R}}^{a \, \sdag} ~.
\end{eqnarray*}

In section~\ref{s4}, dealing with quaternionic matrices 
we shall distinguish between (right) $\mathbb{R}$-linear
quaternionic groups,
\[ \mathsf{GL}\left(N,\mathbb{H}^{\L} \otimes \mathbb{H}^{\R} \right)~,\] 
and $\mathbb{C}$-linear quaternionic groups, 
\[ \mathsf{GL}\left(N,\mathbb{H}^{\L} \otimes \mathbb{C}^{\R} \right)~.\] 
For a clear and complete discussion of standard quaternionic groups, 
\[ \mathsf{GL}\left(N,\mathbb{H}^{\L} \right)~,\] 
the reader is referred to Gilmore's book~\cite{GIL}. The use of
left/right operators give new opportunities in Quaternionic Group Theory. 
Let us observe as follows: The so-called {\em symplectic} 
complex representation of a quaternion (state) 
$q$
\[ q=c_{\1} + j c_{\2}~~~~~~~(c_{\1 ,\2} \in {\mathbb{C}})~,\]  
by a complex column matrix, is
\begin{equation}
\label{sym}
q ~\leftrightarrow ~
\left( \begin{array}{c} c_{\1}\\ c_{\2} \end{array} \right)~. 
\end{equation}
The operator representation of $L_i$, $L_j$ and $L_k$ consistent with the 
above identification
\begin{equation}
\label{ident}
   L_i \leftrightarrow \left( \begin{array}{cc} i & 0\\ 0 & \- i
                              \end{array} \right)=i\sigma_{\3}~,~~~~~
   L_j \leftrightarrow \left( \begin{array}{cc} 0 & \- 1\\ 1 & 0
                              \end{array} \right)= \- i\sigma_{\2}~,~~~~~
   L_k \leftrightarrow \left( \begin{array}{cc} 0 &  \- i\\ \- i & 0
                              \end{array} \right)= \- i\sigma_{\1}~,
\end{equation}
has been known since the discovery of quaternions. It permits any 
quaternionic number or matrix to be translated into a complex matrix, 
{\em but not necessarily vice-versa}. Eight real numbers are required to define
the most general $\od$ complex matrix but only four are needed 
to define
the most general quaternion. In fact since every (non-zero) quaternion has an 
inverse, only a subclass of invertible $\od$ complex matrices are
identifiable with quaternions. $\mathbb{C}$-linear quaternionic operators 
complete the translation~\cite{EVEN}. 
The right quaternionic imaginary unit
\[ R_i ~\leftrightarrow ~ \left( \begin{array}{cc} i & 0\\ 0 & i
                                       \end{array} \right)~,\]
adds four additional degrees of freedom, obtained by matrix multiplication
of the corresponding matrices,
\[ R_i~,~L_i R_i~,~L_j R_i~,~L_k R_i~,\] 
and so we have a set of rules for translating from any $\od$ complex 
matrices to $\mathbb{C}$-linear operators, ${\cal O}_{\mathbb{R}}$. 
This opens new possibilities for quaternionic numbers, see for example 
the one-dimensional version of the Glashow group~\cite{QET}. 
Obviously, this translation does not apply to odd-dimensional complex 
matrices~\cite{ODD}.

%%%%%%%%%%%%%%%%%%%%%%%%%%%%%%%%%%%%%%%%%%%%%%%%%%%%%%%%%%%%%%%%%%%%%%%%%%%%%%%
%                                SECTION IV
%%%%%%%%%%%%%%%%%%%%%%%%%%%%%%%%%%%%%%%%%%%%%%%%%%%%%%%%%%%%%%%%%%%%%%%%%%%%%%%

\section{Quaternionic Groups}
\label{s4}

Every set of basis vectors in $V_{\N}$ a vector space, can be 
related to every other coordinate system by an $\on$ non singular 
matrix. The $\on$ matrix groups involved in changing bases in the 
vector spaces $\mathbb{R}_{\N}$, $\mathbb{C}_{\N}$ and 
$\mathbb{H}^{\L}_{\N}$ are 
called {\em general linear groups} of $\on$ matrices over the reals, 
complex and quaternions
\begin{center}
\begin{tabular}{ccccc}
          & &           & & \\
$\mathsf{GL}\left(N,\mathbb{R} \right)$
          & ~~~~~$\rightarrow$~~~~~ & 
$\mathsf{GL}\left(N, \mathbb{C} \right)$ 
          & ~~~~~$\rightarrow$~~~~~ 
          & $\mathsf{GL}\left(N,\mathbb{H}^{\L} \right)$\\
          & &           & & \\
          & &           & & $\downarrow$  \\
          & &           & & \\
          & & $\mathsf{GL}\left(2N, \mathbb{C} \right)$            
          & ~~~~~$\leftrightarrow$~~~~~ & 
$\mathsf{GL}\left(N,\mathbb{H}^{\L} \otimes \mathbb{C}^{\R} \right)$\\
          & &           & & \\
          & &           & & $\downarrow$  \\
          & &           & & \\
$\mathsf{GL}\left(4N, \mathbb{R} \right)$
          & &  $\leftrightarrow$         & & 
$\mathsf{GL}\left(N,\mathbb{H}^{\L} \otimes \mathbb{H}^{\R} \right)$~.\\
          & &           & & 
\end{tabular}
\end{center}
Before discussing the groups 
$\mathsf{GL}\left(N,\mathbb{H}^{\L} \otimes \mathbb{H}^{\R} \right)$ 
and
$\mathsf{GL}\left(N,\mathbb{H}^{\L} \otimes \mathbb{C}^{\R} \right)$, we 
introduce a new definition of transpose for quaternionic matrices which will 
allow us to overcome the difficulties due to the non-commutative nature 
of the quaternionic field 
(our definition, applying to 
standard quaternions, will be extended to complex and real linear 
quaternions).

\subsection{Quaternionic Transpose Definition}

The customary convention of defining the transpose $M^t$ of the matrix $M$ 
is
\begin{equation} 
\label{st}
(M^t)_{rs} = M_{sr} ~.
\end{equation}
In general, however, for two quaternionic matrices $M$ and $N$ one has
\[ (MN)^t \neq N^t M^t ~,\]
whereas this statement hold as an equality for complex matrices. For example,
the usual definition~(\ref{st}) implies
\begin{equation}
\left[ \left( \begin{array}{cc} q_{\1} & q_{\2} \end{array} \right)
\left( \begin{array}{c} p_{\1} \\ p_{\2} \end{array} \right)
\right]^t = \left( q_{\1} p_{\1} + q_{\2} p_{\2} \right)^t= 
q_{\1} p_{\1} + q_{\2} p_{\2}~,
\end{equation}
and
\begin{equation}
\left( \begin{array}{c} p_{\1} \\ p_{\2} \end{array} \right)^t
\left( 
\begin{array}{cc} q_{\1} & q_{\2} \end{array} 
\right)^t =
p_{\1}^t q_{\1}^t + p_{\2}^t q_{\2}^t = 
p_{\1} q_{\1} + p_{\2} q_{\2}~,
\end{equation}
which are equal only if we use a commutative states. How can we define 
orthogonal quaternionic states and orthogonal groups?  
By looking at the previous example, we see that the 
problem arises in the different position of factors $q_{\1 ,\2}$ and
$p_{\1 ,\2}$. The solution is very simple once
seen. It is possible to give a quaternionic transpose which reverses the 
order of factors. We have three (equivalent) possibilities to define
$q^t$, namely
\begin{equation}
x_{\0} - i x_{\1} +j x_{\2} + k x_{\3}~,~~
x_{\0} + i x_{\1} -j x_{\2} + k x_{\3}~,~~
x_{\0} + i x_{\1} +j x_{\2} - k x_{\3}~.
\end{equation}
We choose
\begin{equation}
\label{tras}
q^t = x_{\0} + i x_{\1} -j x_{\2} + k x_{\3}~,
\end{equation}
which {\em goes back} to the usual definition for complex numbers, 
$c^t=c \in \mathbb{C}(1,i)$. In this way, the transpose of 
a product of two quaternions $q$ and $p$ is the 
product of the transpose quaternions in reverse order
\[ (qp)^t=p^tq^t~.\]
The proof is straightforward if we recognize the following relation 
between transpose $q^t$  and conjugate  $q^{\sdag}$,
\[ q^t= \-j q^{\sdag}j~.\]
What happens for left/right  quaternionic operators? Observing that
for quaternionic states
\[ 
i^t=i~,~~~ j^t= \- j~,~~~ k^t=k~,
\]
the natural generalization for left/right quaternionic operators is 
\[ 
L_i^t=L_i~,~~~ L_j^t= \- L_j~,~~~ L_k^t=L_k~~~\Rightarrow~~~
L_{\mu}^t = \- L_j L_{\mu}^{\sdag} L_j~,
\]
and
\[ 
R_i^t=R_i~,~~~ R_j^t= \- R_j~,~~~ R_k^t=R_k~~~\Rightarrow~~~
R_{\mu}^t = \- R_j R_{\mu}^{\sdag} R_j~.
\]
One-dimensional $\mathbb{R}$-linear 
transpose operators read
\begin{eqnarray}
\label{tro} 
{\cal O}_{\mathbb{R}}^t & = & a^{\mu \nu} L_{\mu}^t \otimes R^t_{\nu} 
\nonumber \\
                         & = & L_j R_j {\cal O}_{\mathbb{R}}^{\sdag} R_j L_j
~~~\in \mathbb{H}^{\L} \otimes \mathbb{H}^{\R}~.
\end{eqnarray}
Thus, in the quaternionic world the transpose 
$M^t$ of the matrix $M \in 
\mathsf{GL}\left(N,\mathbb{H}^{\L} \otimes \mathbb{H}^{\R} \right)$, defined 
by  
\[
(M^t)_{rs} = M^{~~t}_{sr} ~~~\in \mathbb{H}^{\L} \otimes \mathbb{H}^{\R}~,
\]
can be written as
\begin{equation}
\label{m2tras}
 M^t = L_j R_j  M^{\sdag} R_j L_j  = R_j L_j  M^{\sdag} L_j R_j~,
\end{equation}
where 
\[
(M^{\sdag})_{rs} = M^{~~\sdag}_{sr} 
~~~\in \mathbb{H}^{\L} \otimes \mathbb{H}^{\R}~.
\]
With this new 
definition of quaternionic transpose, the relation
\begin{eqnarray*} 
(MN)^t & = & L_j R_j (MN)^{\sdag} R_j L_j = 
             L_j R_j N^{\sdag} M^{\sdag} R_j L_j\\
       & = & L_j R_j N^{\sdag} R_j  L_j L_j  R_j M^{\sdag} R_j L_j\\ 
       & = & N^{t} M^{t} 
\end{eqnarray*}
also holds for non-commutative numbers. Finally, for $\mathbb{C}$-linear 
operators, Eq.~(\ref{tro}) reduces to
\begin{eqnarray}
\label{tco} 
{\cal O}_{\mathbb{C}}^t & = & a^{\mu n} L_{\mu}^t \otimes R_{n} 
\nonumber \\
                         & = & \- L_j {\cal O}_{\mathbb{C}}^{\sdag} L_j~.
\end{eqnarray}
The fundamental property of reverse ring 
the order of factors for the transpose of quaternionic products
is again preserved.

\subsection{$\mathbb{X}$-Mappings}
 
In discussing the classification of the classical (matrix) groups, it is 
necessary to introduce one additional concept: the {\em metric}. Our matrix
element are ${\cal O}_{\mathbb{X}}$-operators, and so it is appropriate
to adopting the metric function, ${\cal M}_{\mathbb{X}}$, mapping of a pair of vectors into a number field $\mathbb{X}$  
\[ {\cal M}_{\mathbb{X}} : 
\mathbb{H}_{\N} \times \mathbb{H}_{\N} \rightarrow \mathbb{X}~,
~~~~~{\cal M}_{\mathbb{X}}(\Psi , \Phi) = (\Psi ,  
\Phi)_{\mathbb{X}}~,\]
with $\Psi, \Phi \in \mathbb{H}_{\N}$ and $(\Psi ,  \Phi)_{\mathbb{X}} 
\in \mathbb{X}$. Let us now recall the following theorem: 
{\em The subset of transformations of basis in $V_{\N}$ which 
preserves the mathematical structure of a metric forms a subgroup of general 
linear groups}. 
\begin{center}
\begin{tabular}{llll}
                   & ~~~bilinear symmetric      &        
                                             & ~~~{\em orthogonal}\\
Groups preserving  & ~~~bilinear antisymmetric  & ~~~metrics are called 
                                             & ~~~{\em symplectic}\\
                   & ~~~sesquilinear symmetric  &
                                             & ~~~{\em unitary}~. 
\end{tabular}
\end{center}
The previous theorem is valid for all real and complex 
metric-preserving matrix groups. It is also valid for quaternionic groups 
that preserve sesquilinear
metrics, since two quaternions obey $(q_{\1}q_{\2})^{\sdag}=q_{\2}^{\sdag}q_{\1}^{\sdag}$. It is not true for quaternionic matrices and bilinear metrics, since two 
quaternions do not generally commute. Nevertheless, it is still possible to 
associate subgroups of  $\mathsf{GL}\left(N,\mathbb{H}^{\L} \right)$ 
with groups that preserve bilinear metrics.
In the literature this is done in the following way. ``{\sl Each quaternion
in $\mathsf{GL}\left(N,\mathbb{H}^{\L} \right)$ is replaced by the 
corresponding $\od$ complex matrix 
using the translation rules~(\ref{ident}). 
The subset of matrices in this complex $\ond$
matrix representation of $\mathsf{GL}\left(N,\mathbb{H}^{\L} \right)$ 
that leaves invariant a bilinear metric 
forms a group, since the theorem is valid for bilinear metrics on complex
linear vector spaces. We can associate an $\on$ quaternion-valued 
matrix with each $\ond$ complex-valued matrix in the resulting groups
that preserve bilinear metrics in the space $\mathbb{C}_{\d \times \d}$, 
which is a representation for the space $\mathbb{H}^{L}_{\N \times \N}$}'' 
- {\em Gilmore}~\cite{GIL}. 

Once we write our complex matrix, we can trivially obtain the generators of 
complex orthogonal groups in a standard manner and then we can translate 
back into quaternionic language. But this is {\em surely a laborious 
procedure}. Defining an appropriate transpose for quaternionic 
numbers~(\ref{tras}), we can overcome the just-cited difficulty. Besides, 
using the symplectic representation~(\ref{sym}), the most general 
transformation (on quaternionic states) will be necessarily represented by 
$\mathbb{C}$-linear quaternionic operators, ${\cal O}_{\mathbb{C}}$, 
and for the invariant 
metric we have to require a ``complex'' projection, 
${\cal M}_{\mathbb{C}} : \mathbb{H} \times \mathbb{H} 
\rightarrow \mathbb{C}$~, 
\[ {\cal M}_{\mathbb{C}}(q^t q) = (q^t q)_ {\mathbb{C}} = 
[(c_{\1} - j c_{\2}^*)(c_{\1} + j c_{\2})]_{\mathbb{C}} = 
c_{\1}^{\2} + c_{\2}^{\2}~.\] 
We wish to emphasize that the introduction of the imaginary unit $R_i$
in complex linear quaternionic operators
\[ (R_i)^{\sdag} = \- R_i~,\]
necessarily implies a complex inner product. The ``new'' imaginary units 
$R_i$ represents an anti-hermitian operator, and so it must verify
\begin{center}
\begin{tabular}{ccc} 
$\int~( R_i \psi)^{\sdag}\varphi$ & ~~~$=$~~~ &  $\- \int~ \psi^{\sdag} R_i
\varphi$\\
         $\downarrow$        &           &     $\downarrow$\\
$\int~(\psi i)^{\sdag}\varphi$ & ~~~$=$~~~ &  
$\- \int~ \psi^{\sdag} \varphi i~.$
\end{tabular}
\end{center}
The previous relation is true only if we adopt a {\em complex projection} 
\[ \int_{\mathbb{C}} \, \equiv \, \frac{1-L_i R_i}{2} \, \int ~,\]
for the inner products
\[ \int_{\mathbb{C}}~(R_i \psi)^{\sdag}\varphi  = \- i \int_{\mathbb{C}}~ 
\psi^{\sdag} \varphi =
\- \int_{\mathbb{C}}~ \psi^{\sdag} \varphi i =
\- \int~ \psi^{\sdag} R_i \varphi
~.\]
Obviously, for $\mathbb{R}$-linear operators
\[ (\vec{R})^{\sdag} = \- \vec{R}~,\]
implies real inner products,
\[ 
{\cal M}_{\mathbb{R}} : \mathbb{H} \times \mathbb{H} 
\rightarrow \mathbb{R}~.
\] 

\subsection{One-Dimensional Quaternionic Groups}

The generators of the unitary and  orthogonal groups 
satisfy the following constraints
\begin{center}
\begin{tabular}{ll}
Groups~:~~~~~~~~~~~~~~ & Generators~:\\
Unitary               & $A+A^{\sdag}=0$ ,\\
Orthogonal            & $A+A^t=0$~.
\end{tabular}
\end{center}
For one-dimensional quaternionic groups, we find  
\begin{center}
\begin{tabular}{|l|l|}
\hline
~{\small Groups}~~~~~~~~~~~~~~  & ~~{\small Generators}\\
\hline
~$\mathsf{U}(1,\mathbb{H}^{\L})$               & ~~$\vec{L}$\\
~$\mathsf{U}(1,\mathbb{H}^{\L} \otimes \mathbb{C}^{\R})$ 
& ~~$\vec{L} \, ,~ R_i$ \\
~$\mathsf{U}(1,\mathbb{H}^{\L} \otimes \mathbb{H}^{\R})$  & 
~~$\vec{L} \, ,~ \vec{R}$\\
                             \hline
~$\mathsf{O}(1, \mathbb{H}^{\L})$      & ~~$L_j$ \\
~$\mathsf{O}(1, \mathbb{H}^{\L} \otimes \mathbb{C}^{\R})$ & 
~~$L_j \, ,~ L_j R_i $ \\
~$\mathsf{O}(1, \mathbb{H}^{\L} \otimes \mathbb{H}^{\R})$ &
 ~~$L_{j}\, ,~L_j R_{i} \, , ~L_j R_{k} \, , ~
R_{j} \, ,~ L_{i} R_j \, , ~L_{k} R_j $\\
\hline
\end{tabular}
\end{center}
At this point, we make a number of observations:

1. - The difference between orthogonal and unitary groups is manifest for 
complex linear quaternionic groups because of the different numbers of 
generators. 

2. - Orthogonal and unitary real linear quaternionic groups have the same 
number of generators.

3. - The real groups $\mathsf{U}(N_+,N_-)$ and 
$\mathsf{O}(N_+,N_-)$ are identical
(there is no difference between bilinear and sesquilinear metrics in a real 
vector space) and this suggest a possible link between 
$\mathsf{U}(N,\mathbb{H}^{\L} \otimes \mathbb{H}^{\R})$ and 
$\mathsf{O}(N,\mathbb{H}^{\L} \otimes \mathbb{H}^{\R})$.

4. - For real linear quaternionic groups, the invariant metric requires a 
``real'' projection (note that $\vec{R}$ represent anti-hermitian 
operators only for real inner products).
Let us show the ``real'' invariant metric for 
$\mathsf{U}(1,\mathbb{H}^{\L} \otimes \mathbb{H}^{\R})$
and $\mathsf{O}(1,\mathbb{H}^{\L} \otimes \mathbb{H}^{\R})$, 
\begin{eqnarray*} 
(x^{\sdag} y)_{\mathbb{R}} &  = &  
[(x_{\0} -ix_{\1}-jx_{\2}-kx_{\3})(y_{\0} +iy_{\1}+jy_{\2}+ky_{\3})]_{\mathbb{R}}=
x_{\0} y_{\0} + x_{\1} y_{\1} + x_{\2} y_{\2} +x_{\3} y_{\3} ~,\\
(x^t y)_{\mathbb{R}} &  = &  
[(x_{\0} +ix_{\1}-jx_{\2}+kx_{\3})(y_{\0} +iy_{\1}+jy_{\2}+ky_{\3})]_{\mathbb{R}}=
x_{\0} y_{\0} - x_{\1} y_{\1} + x_{\2} y_{\2} - x_{\3} y_{\3} ~.
\end{eqnarray*}
We can immediately recognize the invariant metric of 
$\mathsf{O}(4)$ and $\mathsf{O}(2,2)$.
To complete the analogy between one-dimensional real linear quaternionic 
operators and $4$-dimensional real matrices, we observe that besides the
subgroups related to the ${\sdag}$-conjugation (where the sign of all three 
imaginary units is changed)
and the $t$-conjugation (where only $j\rightarrow \-j$), we can define a 
new subgroup which leaves invariant the following real metric
\[ (x^{\sdag} g y)_{\mathbb{R}}~,\]
where
\[ g = \- \f \, L_{\mu} R_{\mu} = 
\- \f \, \left( \u + \vec{L} \cdot \vec{R} \right)~.\]
Explicitly,
\[ (x^{\sdag} g y)_{\mathbb{R}} = 
[(x_{\0} -ix_{\1}-jx_{\2}-kx_{\3})(y_{\0} -iy_{\1}-jy_{\2}-ky_{\3})]_{\mathbb{R}}=
x_{\0} y_{\0} - x_{\1} y_{\1} - x_{\2} y_{\2} -x_{\3} y_{\3} ~.\]
The new subgroup,
\[ 
\tilde{\mathsf{O}}(1,\mathbb{H}^{\L} \otimes \mathbb{H}^{\R})~,~~~~~
g A+A^{\sdag} g=0~,
\]  
represents the one-dimensional quaternionic counterpart of the Lorentz 
group $\mathsf{O}(1,3)$~\cite{SR}. 

The classical groups which occupy a central place in group representation
theory and have many applications in various branches of Mathematics and
Physics are the unitary, special unitary, orthogonal, and symplectic groups.
In order to define special groups, we must define an appropriate
trace for our matrices. In fact, for non-commutative numbers the trace of the
product of two numbers is not the trace of the product with reversed factors.
With complex linear quaternions we have the possibility to give a new 
definition of ``complex'' trace ($Tr$) by
\begin{equation}
\label{ctrace}
Tr~{\cal O}_{\mathbb{C}}^{a} = 
Tr~a^{\mu n} L_{\mu} \otimes R_{n} = 
a^{\t} + i a^{\w}~.
\end{equation}
Such a definition implies that for any two complex linear quaternionic 
operators ${\cal O}_{\mathbb{C}}^{a}$ and ${\cal O}_{\mathbb{C}}^{b}$
\[ Tr~({\cal O}_{\mathbb{C}}^{a} {\cal O}_{\mathbb{C}}^{b}) = 
Tr~({\cal O}_{\mathbb{C}}^{b} {\cal O}_{\mathbb{C}}^{a} )~. \]
For real linear quaternions we need to use the standard definition of ``real''
trace ($tr$)
\begin{equation}
\label{rtrace}
tr~{\cal O}_{\mathbb{R}}^a = a^{\t}~,
\end{equation}
since the previous ``complex'' definition~(\ref{ctrace}) gives
\[ Tr~({\cal O}_{\mathbb{R}}^{a} {\cal O}_{\mathbb{R}}^{b}) \neq 
Tr~({\cal O}_{\mathbb{R}}^{b} {\cal O}_{\mathbb{R}}^{a})~. \]
For example, for
\[ {\cal O}_{\mathbb{R}}^{a} = R_j~~~\mbox{and}~~~ 
   {\cal O}_{\mathbb{R}}^{b} = R_k~,
\]
we find
\begin{eqnarray*}
Tr~({\cal O}_{\mathbb{R}}^{a} {\cal O}_{\mathbb{R}}^{b}) & = &
Tr~(R_j R_k) = \- Tr~(R_i) = \- 1 ~,\\
Tr~({\cal O}_{\mathbb{R}}^{b} {\cal O}_{\mathbb{R}}^{a}) & = & 
Tr~(R_k R_j) = \+ Tr~(R_i) = \+ 1 ~.
\end{eqnarray*}

\subsection{$\mathbf{N}$-Dimensional Quaternionic Groups}

We recall that the generators of the unitary, special unitary, 
orthogonal groups must satisfy the following conditions~\cite{COR}
\begin{center}
\begin{tabular}{lll}
$\mathsf{U}(N)$   & ~~~~~~~$A+A^{\sdag}=0$         , &                 \\
$\mathsf{SU}(N)$  & ~~~~~~~$A+A^{\sdag}=0$         , &~~~ $Tr~A=0$ ,   \\
$\mathsf{O}(N)$   & ~~~~~~~$A+A^{t}=0$             . &               
\end{tabular}
\end{center}
These conditions also apply for quaternionic groups. For complex symplectic 
groups we find
\begin{center}
\begin{tabular}{lll}
$\mathsf{Sp}(2N)$ & ~~~~~~~${\cal J}A+A^{t} {\cal J}=0$  , & 
\end{tabular}
\end{center}
where
\[ {\cal J} = \left( \begin{array}{cc} 
\mathbf{0}_{\N \times \N} & \mathbf{1}_{\N \times \N}\\
\- \mathbf{1}_{\N \times \N} & \mathbf{0}_{\N \times \N}
              \end{array} \right)~.\]
Working with quaternionic numbers, we can construct a group preserving a 
non-singular antisymmetric metric, for $N$ odd as well 
as $N$ even. Thus for quaternionic symplectic groups we have
\begin{center}
\begin{tabular}{lll}
$\mathsf{Sp}(N)$ & ~~~~~~~${\cal J}A+A^{t} {\cal J}=0$  , & 
\end{tabular}
\end{center}
with
\[ {\cal J}_{\d \times \d} = \left( \begin{array}{cc} 
\mathbf{0}_{\N \times \N} & \mathbf{1}_{\N \times \N}\\
\- \mathbf{1}_{\N \times \N} & \mathbf{0}_{\N \times \N}
              \end{array} \right)~, ~~~~~~~ 
{\cal J}_{( \d +1)\times ( \d +1)} = \left( \begin{array}{ccc} 
\mathbf{0}_{\N \times \N} & \mathbf{0}_{\N \times \1}  & 
\mathbf{1}_{\N \times \N}\\
\mathbf{0}_{\1 \times \N} & L_j  & \mathbf{0}_{\1 \times \N}\\
\- \mathbf{1}_{\N \times \N} & \mathbf{0}_{\N \times \1} & 
\mathbf{0}_{\N \times \N}
              \end{array} \right)~.
\]

The generators of one-dimensional groups with complex and real linear
quaternions are given in the following table
\begin{center}
\begin{tabular}{|c|}
\hline
~~One-dimensional  quaternionic groups~~\\
\hline
\end{tabular}
\end{center} 
\begin{center}
\begin{tabular}{|l|lcl|}
\hline
~{\small Groups} & ~{\small Generators}  & &  \\
\hline
~$\mathsf{U}(1,\mathbb{H}^{\L} \otimes \mathbb{C}^{\R})$   & 
~${\cal A}_{\mathbb{C}} + {{\cal A}_{\mathbb{C}}}^{\dag}=0$           
                    & :
                    & $\vec{L} \, ,~R_i$\\
~$\mathsf{SU}(1,\mathbb{H}^{\L} \otimes \mathbb{C}^{\R})$  & 
~${\cal A}_{\mathbb{C}} + {{\cal A}_{\mathbb{C}}}^{\dag}=0$,~ 
     $Tr~{\cal A}_{\mathbb{C}}=0$      
                    & :
                    & $\vec{L}$ \\
~$\mathsf{O}(1,\mathbb{H}^{\L} \otimes \mathbb{C}^{\R})$   & 
~${\cal A}_{\mathbb{C}} + {{\cal A}_{\mathbb{C}}}^t=0$ 
                    & :
                    & $L_j \, ,~L_j R_i$  \\
~$\mathsf{Sp}(1,\mathbb{H}^{\L} \otimes \mathbb{C}^{\R})$  
         & ~$L_j {\cal A}_{\mathbb{C}} + {{\cal A}_{\mathbb{C}}}^t L_j=0$ 
         & :
         & $\vec{L} \, , ~\vec{L} R_i$ \\
\hline
~$\mathsf{U}(1,\mathbb{H}^{\L} \otimes \mathbb{H}^{\R})$   & 
~${\cal A}_{\mathbb{R}} + {{\cal A}_{\mathbb{R}}}^{\dag}=0$ 
                    & :
                    & $\vec{L} \, , ~ \vec{R}$ \\
~$\mathsf{O}(1,\mathbb{H}^{\L} \otimes \mathbb{H}^{\R})$   & 
~${\cal A}_{\mathbb{R}} + {{\cal A}_{\mathbb{R}}}^t=0$ 
                    & :
                    & $L_j \, , ~ L_j R_{i} \, , ~L_j R_{k} \, , ~
                             R_j \, , ~ L_i R_{j} \, , ~L_k R_{j}    $~ \\ 
~$\tilde{\mathsf{O}}(1,\mathbb{H}^{\L} \otimes \mathbb{H}^{\R})$   & 
~$g{\cal A}_{\mathbb{R}} + {{\cal A}_{\mathbb{R}}}^{\sdag} g=0$ 
          & :
          & $\vec{L} - \vec{R} \, , ~ \vec{L} \times \vec{R}$\\
~$\mathsf{Sp}(1,\mathbb{H}^{\L} \otimes \mathbb{H}^{\R})$  & 
~$L_j {\cal A}_{\mathbb{R}} + {{\cal A}_{\mathbb{R}}}^t L_j=0$ 
                    & :
                    & $\vec{L} \, , ~ \vec{L} R_i \, , ~
                             \vec{L} R_k \, , ~ R_j$~ \\ 
\hline
\end{tabular}
\end{center}

We conclude our classification of quaternionic groups  
giving the general formulas for counting the 
generators of generic $N$-dimensional groups as function of $N$.
\begin{center}
\begin{tabular}{|c|}
\hline
~~N-dimensional quaternionic groups~~\\
\hline
\end{tabular}
\end{center}
\begin{center}
\begin{tabular}{|lclcr|}
\hline
~$\mathsf{U}(N,{\mathbb{H}}^{\L})$ & ~$\leftrightarrow$~ & 
$\mathsf{USp}(2N,\mathbb{C})$ & ~~~~~~~~~~~ & $N(2N+1)$ ~\\
~$\mathsf{U}(N,\mathbb{H}^{\L} \otimes \mathbb{C}^{\R})$ & 
~$\leftrightarrow$~ & $\mathsf{U}(2N,\mathbb{C})$ & ~~~~~~~~~~~ & $4N^{\2}$~\\
~$\mathsf{U}(N,\mathbb{H}_{\L} \otimes \mathbb{H}^{\R})$ &
~$\leftrightarrow$~ & $\mathsf{O}(4N)$ & ~~~~~~~~~~~ & $2N(4N-1)$ ~\\
\hline
~$\mathsf{SU}(N,{\mathbb{H}}^{\L})$  & ~$\equiv$~ & 
$\mathsf{U}(N,{\mathbb{H}}^{\L})$ & ~~~~~~~~~~~ & \\
~$\mathsf{SU}(N,\mathbb{H}^{\L} \otimes \mathbb{C}^{\R})$ & 
~$\leftrightarrow$~ & $\mathsf{SU}(2N,\mathbb{C})$ & ~~~~~~~~~~~ & 
$4N^{\2}-1$~\\
~$\mathsf{SU}(N,\mathbb{H}^{\L} \otimes \mathbb{H}^{\R})$ & ~$\equiv$~ & 
$\mathsf{U}(N,\mathbb{H}^{\L} \otimes \mathbb{H}^{\R})$ & ~~~~~~~~~~~ & \\
\hline
~$\mathsf{O}(N,{\mathbb{H}}^{\L})$  & ~$\leftrightarrow$~ & 
$\mathsf{SO}^{*}(2N,\mathbb{C})$ & ~~~~~~~~~~~ & $N(2N-1)$~\\
~$\mathsf{O}(N,\mathbb{H}^{\L} \otimes \mathbb{C}^{\R})$ & 
~$\leftrightarrow$~ & 
$\mathsf{O}(2N,\mathbb{C})$ & ~~~~~~~~~~~ & $2N(2N-1)$~\\
~$\mathsf{O}(N,\mathbb{H}^{\L} \otimes \mathbb{H}^{\R})$ & 
~$\leftrightarrow$~ & 
$\mathsf{O}(2N_{\+},2N_{\-})$ & ~~~~~~~~~~~ & $2N(4N-1)$~\\
~$\tilde{\mathsf{O}}(N,\mathbb{H}^{\L} \otimes \mathbb{H}^{\R})$ 
                     & ~$\leftrightarrow$~ & $\mathsf{O}(3N_{\+},N_{\-})$ & 
~~~~~~~~~~~ & $2N(4N-1)$~\\
\hline
~$\mathsf{Sp}(N,{\mathbb{H}}^{\L})$  & ~$\leftrightarrow$~ & 
$\mathsf{USp}(2N,\mathbb{C})$ & ~~~~~~~~~~~ & $N(2N+1)$~\\
~$\mathsf{Sp}(N,\mathbb{H}^{\L} \otimes \mathbb{C}^{\R})$ & 
~$\leftrightarrow$~ & $\mathsf{Sp}(2N,\mathbb{C})$ & ~~~~~~~~~~~ & 
$2N(2N+1)$~\\
~$\mathsf{Sp}(N,\mathbb{H}^{\L} \otimes \mathbb{H}^{\R})$ & 
~$\leftrightarrow$~ & $\mathsf{Sp}(4N,\mathbb{R})$ & ~~~~~~~~~~~ & 
$2N(4N+1)$~\\
\hline
\end{tabular}
\end{center}

%%%%%%%%%%%%%%%%%%%%%%%%%%%%%%%%%%%%%%%%%%%%%%%%%%%%%%%%%%%%%%%%%%%%%%%%%%%%%%%
%                                SECTION V
%%%%%%%%%%%%%%%%%%%%%%%%%%%%%%%%%%%%%%%%%%%%%%%%%%%%%%%%%%%%%%%%%%%%%%%%%%%%%%%

\section{Physical Applications}
\label{s5}

In the last years the left/right action of the quaternionic numbers, 
expressed by left/right operators, 
$\mathcal{O}_{\mathbb{R}}$, $\mathcal{O}_{\mathbb{C}}$, 
$\mathcal{O}_{\mathbb{H}}$,  
has been very useful in overcoming difficulties owing 
to the non-commutativity of quaternions. Among the successful applications 
of left/right operators we mention the 
one-dimensional quaternionic formulation of Lorentz boosts. 

\subsection{Special Relativity}
The Lorentz group, $\mathsf{O}(3,1)$, is  characterized by six parameters, 
three for rotations and three for boosts. 
Corresponding to these six parameters there are six 
generators. The anti-hermitian generators associated to spatial rotations
and the hermitian boost generators satisfy the following commutation relations
\begin{eqnarray}
\mathcal{A}_x = [ \mathcal{A}_y , \mathcal{A}_z ]~,~~~ 
\mathcal{A}_x = [ \mathcal{B}_z , \mathcal{B}_y ]~,~~~
\mathcal{B}_x = [ \mathcal{A}_y , \mathcal{B}_z ]
              = [ \mathcal{B}_y , \mathcal{A}_z ]~, \nonumber\\
\mathcal{A}_y = [ \mathcal{A}_z , \mathcal{A}_x ]~,~~~ 
\mathcal{A}_y = [ \mathcal{B}_x , \mathcal{B}_z ]~,~~~ 
\mathcal{B}_y = [ \mathcal{A}_z , \mathcal{B}_x ]
              = [ \mathcal{B}_z , \mathcal{A}_x ] ~,\\ 
\mathcal{A}_z = [ \mathcal{A}_x , \mathcal{A}_y ]~,~~~
\mathcal{A}_z = [ \mathcal{B}_y , \mathcal{B}_x ]~,~~~ 
\mathcal{B}_z = [ \mathcal{A}_x , \mathcal{B}_y ]
              = [ \mathcal{B}_x , \mathcal{A}_y ]~, \, \nonumber
\end{eqnarray}
The idea of combining left and right imaginary units gives the possibility
to obtain a one-dimensional quaternionic representation for boost generators
by $\mathbb{R}$-linear quaternionic operators~\cite{SR}
\begin{center}
\begin{tabular}{ll}
Boost generators: & $\frac{1}{2} \, {\vec L} \times {\vec R}~,$\\
  &  \\
Rotation generators: & $\frac{1}{2} \, \left( {\vec L} - {\vec R} \right)~.$ 
\end{tabular}
\end{center}
The four real quantities which identify the space-time point $(ct,x,y,z)$
are represented by the quaternion
\[ q=ct + i x + j y + k z~.\]
The one-dimensional group 
$\tilde{\mathsf{O}}(1, \mathbb{H}^{\L} \otimes \mathbb{H}^{\R})$
represents the quaternionic counterpart of the four-dimensional Lorentz
group $\mathsf{O}(3,1)$. Infinitesimal rotations about the $x$-axis and 
boosts $(ct,x)$ are respectively given by
\begin{eqnarray}
\tilde{q}_{r} & = & 
\left[ 1 + \stx \left( L_i - R_i \right) \right] \, q \nonumber \\
          & = & 
ct + ix + j (y - \theta_x z) + k (z + \theta_x y)~,\\
\tilde{q}_{b} & = & 
\left[ 1 + \svx \left( L_j R_k  -L_k R_j \right) \right] \, q 
\nonumber \\
          & = & 
ct + \varphi_x x + i ( x + \varphi_x ct) + j y + k z~.
\end{eqnarray}

In analogy to the connection between the rotation group
$\mathsf{O}(3)$ and the special unitary group $\mathsf{SU}(2)$, 
there is a natural 
correspondence between the Lorentz group $\mathsf{O}(3,1)$ and the special 
linear group $\mathsf{SL}(2,\mathbb{C})$. 
The use of left/right quaternionic imaginary units gives the possibility 
to extend such connections to ``quaternionic'' group,
\begin{center}
\begin{tabular}{lclcl}
$\mathsf{O}(3)$     & $~\sim~$ & $\mathsf{SU}(2)$ & $~\leftrightarrow~$ & 
$\mathsf{U}(1,\mathbb{H}^{\L})$~,\\
$\mathsf{O}(3,1)$   & $~\sim~$ & $\mathsf{SL}(2,\mathbb{C})$  
& $~\leftrightarrow~$ & 
$\mathsf{SL}(1,\mathbb{H}^{\L} \otimes \mathbb{C}^{\R})$~.
\end{tabular}
\end{center}
In fact, combining left quaternionic imaginary units, $\vec{L}$, with right 
``complex'' imaginary unit, $R_i$, we obtain the following one-dimensional 
representation for rotation and boost generators
\begin{equation}
\label{ql1}
\mathcal{A}_x = L_i/2~,~~~
\mathcal{A}_y = L_j/2~,~~~
\mathcal{A}_z = L_k/2~~~~~\in \mathbb{H}^{\L}~,
\end{equation}
and 
\begin{equation}
\label{ql2}
\mathcal{B}_x = L_i R_i /2~,~~~
\mathcal{B}_y = L_j R_i/2~,~~~
\mathcal{B}_z = L_k R_i/2~~~~~\in 
\mathbb{H}^{\L} \otimes \mathbb{C}^{\R}~.
\end{equation}
Now, the four real quantities which identify the space-time point are 
represented by the symplectic decomposition of quaternionic spinor states
\begin{equation}
\label{qs}
q = \xi + j \eta~,~~~~~\xi , \eta \in \mathbb{C}(1,i)~,
\end{equation}
where
\[
q (1+i) q^{\sdag} \equiv ct + ix + jy +kz~.
\]
With this identifications, $\mathsf{O}(3,1)$-transformations on
\[
X = \left( \begin{array}{c} ct\\
                            x \\ 
                            y \\
                            z
                       \end{array} \right)~,
\] 
are  equivalent to one-dimensional transformations on 
quaternionic spinors~(\ref{qs}). A detailed discussion is found 
in ref.~\cite{CHI}. 

New possibilities, coming out from the use of $\mathbb{C}$-linear  
quaternionic operators, also appear 
in Quantum Mechanics and Field Theory, e.g. they allow an appropriate
definition of the momentum operator~\cite{DIRAC}, quaternionic version of 
standard relativistic equations~\cite{DIRAC,DKP}, Lagrangian 
formalism~\cite{LAG}, electroweak model~\cite{QET} and grand unification 
theories~\cite{GUT}.

\subsection{Dirac Equation}

In the complex world, the Dirac equation read indifferently as
\[ 
i \partial_t \psi = H \psi ~~~\mbox{or}~~~
   \partial_t \psi i = H \psi ~.
\]
In the quaternionic world there is a clear difference in choosing a left or 
right position for our complex imaginary unit $i$. In fact, 
by requiring norm conservation
\[ \partial_t \int d^{\3} x \, \psi^{\sdag} \psi  = 0~, \]
we find that a left position of the imaginary unit $i$ in the quaternionic
Dirac equation,  
\[ L_i \partial_t \psi \equiv i \partial_t \psi = H \psi~,  \]
gives
\[ 
\partial_t \int d^{\3} x \, \psi^{\sdag} \psi  = 
\int d^{\3} x \, \psi^{\sdag} [ H , i ] \psi~, 
\]
in general $\neq 0$ for quaternionic Hamiltonians. A right position
of the imaginary unit $i$, 
\[ R_i \partial_t \psi \equiv \partial_t \psi i = H \psi~,  \]
ensures the norm conservation. From covariance, by treating time and space in 
the same way, we obtain the following  
``quaternionic'' momentum operator  
\begin{equation}
p^{\mu} \leftrightarrow R_i \partial^{\mu}~~~\Rightarrow~~~
p^{\mu} \psi \leftrightarrow R_i \partial^{\mu} \psi \equiv 
\partial^{\mu} \psi i~.
\end{equation}
Finally, the quaternionic Dirac equation reads
\begin{equation}
\label{dirac1}
R_i \gamma^{\mu} \partial_{\mu} \psi \equiv 
\gamma^{\mu} \partial_{\mu} \psi i = m \psi~,~~~
\left[ \gamma^{\mu} ,  \gamma^{\nu} \right] = 2 g^{\mu \nu}~,
\end{equation}
with
\[ 
\gamma_{\mu} \in \mathcal{M}_{\2} ( \mathbb{H}^{\L} \otimes \mathbb{C}^{\R} )
~~~\mbox{and}~~~ \psi \in \mathbb{H}_{\2}~.
\]

Another fundamental ingredient in the formulation of quaternionic relativistic 
quantum mechanics is represented by the adoption of a 
{\em complex geometry}~\cite{REM},
necessary in order to guarantee  that 
$R_i \vec{\partial}$ be an hermitian operator
\[ 
\int d^{\3} x \, \varphi^{\sdag} R_i \vec{\partial} \psi  = 
\int d^{\3} x \,  ( R_i \vec{\partial} \varphi ) ^{\sdag}  \psi ~.
\]
The previous relation implies
\[ 
\int d^{\3} x \, \varphi^{\sdag} \vec{\partial} \psi i  = 
\- i \int d^{\3} x \,  \vec{\partial} \varphi^{\sdag}  \psi~ 
\mbox{\footnotesize (after integration by parts)~}
= i \int d^{\3} x \, \varphi^{\sdag} \vec{\partial} \psi ~.
\]
The different position of the imaginary unit $i$ forces the use of a 
{\em complex projection}~\cite{REM} for inner products
\begin{equation}
\int d^{\3} x  ~\rightarrow~ \int_{\mathbb{C}} d^{\3} x~.  
\end{equation}

\subsection{Lagrangian Formalism}

The use of the variational principle within quaternionic quantum 
mechanics is non-trivial because of the non commutative nature of 
quaternions. In this subsection, we write the Dirac Lagrangian density 
corresponding to the two-component Dirac equation. This Lagrangian is 
{\em complex projected} as anticipated in previous articles~\cite{LAG}.
The traditional form for the Dirac Lagrangian density is
\begin{equation}
\label{lag}
\mathcal{L} =   i \bar{\psi} \gamma^{\mu} \partial_{\mu} \psi - m 
                \bar{\psi} \psi~.
\end{equation}
The position of the imaginary unit $i$ is purely 
conventional in~(\ref{lag}) but with a quaternionic 
number field we must recognize that the $\partial_{\mu}$ operator is more 
precisely part of the first quantized momentum 
operator $R_i \partial_{\mu}$ and that hence to ensure $\mathcal{L}$ be an   
hermitian quantity we must taken a complex projection of
the kinetic term
\begin{eqnarray*}
\mathcal{L}_{kin}  & = &  
\left(  \bar{\psi} \gamma^{\mu} R_i \partial_{\mu} \psi \right)_{\mathbb{C}}\\
                   & \equiv &
\left(  \bar{\psi} \gamma^{\mu}  \partial_{\mu} 
\psi i \right)_{\mathbb{C}}\\
                   & = & 
\left(  \bar{\psi} \gamma^{\mu} \partial_{\mu} \psi \right)_{\mathbb{C}} 
i~.
\end{eqnarray*}  
The requirement of hermiticity however says nothing about the Dirac mass 
term in eq.~(\ref{lag}). It is here that appeal to the variational 
principle must be made. A variation $\delta \psi$ in $\psi$ cannot  
brought to the extreme right because of the imaginary unit $i$. 
The only consistent procedure is to 
generalize the variational rule that says that $\psi$ and $\bar{\psi}$ 
must be varied {\em independently}. 
We thus apply independent variations to $\psi$ 
($\delta \psi$) and $\psi i$ ($\delta (\psi i)$). 
Similarly for $\delta \bar{\psi}$ and 
$\delta (i \bar{\psi})$. Now, to obtain the desired Dirac equation for $\psi$ 
and its adjoint equation for $\bar{\psi}$, we must modify the mass 
term into
\[
{\cal L}_{mass}= - m \left( \bar{\psi} \psi \right)_{\mathbb{C}}~. 
\]
The final result for $\cal L$ is
\begin{equation}
\label{lag2}
\mathcal{L} =   \left( \bar{\psi} \gamma^{\mu} R_i \partial_{\mu} \psi - m 
                \bar{\psi} \psi \right)_{\mathbb{C}}~.
\end{equation}

\subsection{Electroweak Models}

Let us now eaxamine the  fermion/quark sector of Salam-Weinberg 
model~\cite{SW}. The first family is represented by
\[ 
\left( \begin{array}{c} \nu \\ e  \end{array} \right)~~~\mbox{and}~~~
\left( \begin{array}{c} u \\ d  \end{array} \right)~.
\]

In the standard representation~\cite{QET,DIRAC}  
\begin{equation}
\Psi_{\L} = \left( \begin{array}{cc} \nu_{\L} &  u_{\L}\\
                                e_{\L} & d_{\L} 
\end{array} \right)~,~~~
\Psi_{\R} = \left( \begin{array}{cc} \nu_{\R} &  u_{\R}\\
                                e_{\R} & d_{\R} 
\end{array} \right)~~~~\in \mathcal{M}_{\4}(\mathbb{H} )~.
\end{equation}
The massless fermion electroweak Lagrangian 
\begin{equation}
\mathcal{L}_{fermion} = \left(   
\bar{\Psi}_{\L} \gamma^{\mu} R_i \partial_{\mu} \Psi_{\L} +
\bar{\Psi}_{\R} \gamma^{\mu} R_i \partial_{\mu} \Psi_{\R} 
\right)_{\mathbb{C}}~,
\end{equation}
where
\[ 
\gamma_{\mu} \in \mathcal{M}_{\4} ( \mathbb{C}^{\R} )~,
\]
is global invariant under the quaternionic Glashow gauge group~\cite{GLA}
\begin{equation}
\mathsf{U}(1, \mathbb{H}^{\L})_{\L} \otimes 
\mathsf{U}(1, \mathbb{C}^{\R})_{\Y}~.
\end{equation}

In the chiral representation~\cite{CHI} $\nu_{\L \b \R }$, 
$e_{\L \b \R }$, $u_{\L \b \R }$ and  $d_{\L \b \R }$ 
are one-dimensional quaternionic spinors and so can be accommodate in 
\begin{equation}
\Psi_{\L} = \left( \begin{array}{cc} \nu_{\L} &  u_{\L}\\
                                e_{\L} & d_{\L} 
\end{array} \right)~,~~~
\Psi_{\R} = \left( \begin{array}{cc} \nu_{\R} &  u_{\R}\\
                                e_{\R} & d_{\R} 
\end{array} \right)~~~~\in \mathcal{M}_{\2}(\mathbb{H} )~.
\end{equation}
The Lagrangian for the masseless fermion sector 
\begin{equation}
\mathcal{L}_{fermion} = \left(   
\bar{\Psi}_{\L} \gamma^{\mu} R_i \partial_{\mu} \Psi_{\L} +
\bar{\Psi}_{\R} \gamma^{\mu} R_i \partial_{\mu} \Psi_{\R} 
\right)_{\mathbb{C}}~,
\end{equation}
is now global invariant under the ``right complex'' gauge group
\begin{equation}
\mathsf{SU}(2, \mathbb{C}^{\R})_{\L} \otimes 
\mathsf{U}(1, \mathbb{C}^{\R})_{\Y}~.
\end{equation}

%%%%%%%%%%%%%%%%%%%%%%%%%%%%%%%%%%%%%%%%%%%%%%%%%%%%%%%%%%%%%%%%%%%%%%%%%%%%
%%%%%%                CONCLUSIONS
%%%%%%%%%%%%%%%%%%%%%%%%%%%%%%%%%%%%%%%%%%%%%%%%%%%%%%%%%%%%%%%%%%%%%%%%%%%%

\section{Conclusions}
\label{s6}

The more exciting possibility that quaternionic or octonionic equations will
eventually play a significant role in Mathematics and Physics is synonymous, 
for some physicist, with
the advent of a revolution in Physics comparable to that of Quantum Mechanics.

For example, Adler suggested~\cite{ADLP} that the color degree of freedom
postulated in the Harari-Shupe model~\cite{H,S} (where we can think of
quarks and leptons as composites of other more fundamental fermions, preons) 
could be sought in a non-commutative extension of the complex field.
Surely a stimulating idea. Nevertheless, we think that it would be very 
strange if standard Quantum 
Mechanics did not permit a quaternionic or octonionic description other than
in the trivial sense that complex numbers are contained within the quaternions
or octonions. 

In the last few years much progress has been achieved in manipulating
such fields. We quote the quaternionic version of electroweak 
theory~\cite{QET}, where the Glashow group is expressed by the one-dimensional
quaternionic group $\mathsf{U}(1,\mathbb{H}^{\L}) \otimes 
\mathsf{U}(1,\mathbb{C}^{\R})$,  
quaternionic GUTs~\cite{GUT} and Special Relativity, where the Lorentz group 
is represented by $\tilde{\mathsf{O}}(1,\mathbb{H}^{\L} \otimes 
\mathbb{H}^{\R})$.  We also recall new possibilities related to the 
use of octonions in Quantum Mechanics~\cite{OQM}, in particular in writing a 
one-dimensional octonionic Dirac equation. The link between 
octonionic and quaternionic versions of standard Quantum Physics is represented
by the use of a complex geometry~\cite{REM}.

In this paper we observed that beyond the study of matrix groups with 
``simple'' quaternionic elements, $\mathcal{O}_{\mathbb{H}}$, 
one can consider more general groups 
with matrix elements of the form ${\cal O}_{\mathbb{R}}$ and  
${\cal O}_{\mathbb{C}}$.
To the best of our 
knowledge these more general matrix groups have not been studied in the 
literature.  We overcome the problems arising in the definitions of transpose, 
determinant and trace for quaternionic matrices. 
For octonionic fields~\cite{OREP} we must
admit a more complicated situation, yet our discussion can be also proposed  
for non-associative numbers. 

Finally, we hope that this paper emphasizes the possibility of using 
hyper-complex numbers in Mathematics and Physics and could  represent an 
important step towards a complete and clear discussion on 
Hyper-complex Group and Field Theories. 

\section*{Acknowledgements}

The authors wish to thank the Prof.~Dr.~Nir Cohen for helpful comments and
stimulating conversations on quaternionic group theory 
and non-commutative algebras. One of us (GCD) is acknowledges the 
CAPES for financial support. 

%%%%%%%%%%%%%%%%%%%%%%%%%%%%%%%%%%%%%%%%%%%%%%%%%%%%%%%%%%%%%%%%%%%%%%%%%%%%%%%

%%%%%%%%%%%%%%%%%%%%%%%%%%%%%%%%%%%%%%%%%%%%%%%%%%%%%%%%%%%%%%%%%%%%%%%%%%%%%%%
%                                APPENDIX A
%%%%%%%%%%%%%%%%%%%%%%%%%%%%%%%%%%%%%%%%%%%%%%%%%%%%%%%%%%%%%%%%%%%%%%%%%%%%%%%

\section*{Appendix A}

We give the translation rules between quaternionic  
left/right $\mathbb{R}$-linear operators and $\oq$ real matrices:
{\scriptsize
\begin{eqnarray*}
{L_i} \leftrightarrow \left( \begin{array}{cccc} 
0 & \- 1    & 0 & 0 \\ 
1 & 0    & 0 & 0 \\
0 & 0    & 0 & \- 1 \\
0 & 0    & 1 & 0                              
\end{array} \right)~,~~~
{L_j}  \leftrightarrow \left( \begin{array}{cccc} 
0 & 0    &\- 1 & 0 \\ 
0 & 0    & 0 & 1 \\
1 & 0    & 0 & 0 \\
0 & \- 1    & 0 & 0                              
\end{array} \right)~,~~~ 
L_k = L_i L_j~,\\
{R_i}  \leftrightarrow \left( \begin{array}{cccc} 
0 & \- 1    & 0 & 0 \\ 
1 & 0    & 0 & 0 \\
0 & 0    & 0 & 1 \\
0 & 0    & \- 1 & 0 
\end{array} \right)~,~~~
{R_j}  \leftrightarrow \left( \begin{array}{cccc} 
0 & 0    & \- 1 & 0 \\ 
0 & 0    & 0 & \- 1 \\
1 & 0    & 0 & 0 \\
0 & 1    & 0 & 0 
\end{array} \right)~,~~~
R_k = R_j R_i~.
\end{eqnarray*}
}
From these identifications we can obtain the {\em full} translation. 
For example, the matrix counterpart of the operator $L_j R_k$ is soon 
achieved by
{\scriptsize
\begin{eqnarray*}
L_j R_k & ~\leftrightarrow ~ & \left( \begin{array}{cccc} 
0 & 0    & \- 1 & 0 \\ 
0 & 0    & 0 & 1 \\
1 & 0    & 0 & 0 \\
0 & \- 1    & 0 & 0   
\end{array} \right) 
\left[ 
\left( \begin{array}{cccc} 
0 & 0    & \- 1 & 0 \\ 
0 & 0    & 0 & \- 1 \\
1 & 0    & 0 & 0 \\
0 & 1    & 0 & 0 
\end{array} \right)
\left( \begin{array}{cccc} 
0 & \- 1    & 0 & 0 \\ 
1 & 0    & 0 & 0 \\
0 & 0    & 0 & 1 \\
0 & 0    & \- 1 & 0 
\end{array} \right)
\right] = \\
&  & 
\left( \begin{array}{cccc} 
0 & 0    & \- 1 & 0 \\ 
0 & 0    & 0 & 1 \\
1 & 0    & 0 & 0 \\
0 & \- 1    & 0 & 0   
\end{array} \right) 
\left( \begin{array}{cccc} 
0 & 0    & 0 & \- 1 \\ 
0 & 0    & 1 & 0 \\
0 & \- 1    & 0 & 0 \\
1 & 0    & 0 & 0 
\end{array} \right)
=
\left( \begin{array}{cccc} 
0 & 0    & 0 & \- 1 \\ 
0 & 0    & 1 & 0 \\
0 & \- 1    & 0 & 0 \\
1 & 0    & 0 & 0 
\end{array} \right)
\left( \begin{array}{cccc} 
0 & 0    & \- 1 & 0 \\ 
0 & 0    & 0 & 1 \\
1 & 0    & 0 & 0 \\
0 & \- 1    & 0 & 0   
\end{array} \right) 
= \\
 &  & \left( \begin{array}{cccc} 
0 & 1    & 0 & 0 \\ 
1 & 0    & 0 & 0 \\
0 & 0    & 0 & \- 1 \\
0 & 0    & \- 1 & 0   
\end{array} \right)~. 
\end{eqnarray*}
}

%%%%%%%%%%%%%%%%%%%%%%%%%%%%%%%%%%%%%%%%%%%%%%%%%%%%%%%%%%%%%%%%%%%%%%%%%%%%%%%
%                                REFERENCES
%%%%%%%%%%%%%%%%%%%%%%%%%%%%%%%%%%%%%%%%%%%%%%%%%%%%%%%%%%%%%%%%%%%%%%%%%%%%%%%

%%%%%%%%%%%%%%%%%%%%%%%%%%%%%%%%%%%%%%%%%%%%%%%%%%%%%%%%%%%%%%%%%%%%%%%%%%%%%%
\end{document}